\documentclass[aps,prl,twocolumn,superscriptaddress,showpacs]{revtex4}
\usepackage{graphicx}
\newcommand{\figwidth}{3.375in}

\begin{document}

\title{Eigenmode analysis of the susceptibility matrix of the
four-dimensional Edwards--Anderson spin-glass model}  

\author{Koji Hukushima}
\email{hukusima@issp.u-tokyo.ac.jp}
\affiliation{
Institute for Solid State Physics, University~of~Tokyo, 5-1-5
Kashiwa-no-ha, Kashiwa, Chiba 277-8581, Japan} 
\author{Yukito Iba}
\email{iba@ism.ac.jp}
\affiliation{
Department of Prediction and Control, 
The Institute of Statistical Mathematics
4-6-7, Minami-Azabu, Minatoku 106-8569, Tokyo, Japan}

\begin{abstract}
The nature of spin-glass phase of the four-dimensional Edwards-Anderson
Ising model is numerically studied by eigenmode analysis of the 
susceptibility matrix up to the lattice size $10^4$.  
Unlike the preceding results on smaller lattices, our result suggests 
that there exist multiple extensive eigenvalues of the matrix, which
does not contradict replica-symmetry-breaking scenarios.      
The sensitivity of the eigenmodes with respect to a temperature change 
is examined using finite-size-scaling analysis and 
an evidence of anomalous sensitivity is found.
A computational advantage of dual formulation of the eigenmode analysis
in the study of large lattices is also discussed. 
\end{abstract}
\date{July 4, 2002}


\maketitle

\sloppy

An essential problem in the study of randomly frustrated
systems such as spin glass (SG) is how to choose appropriate order
parameters. In these systems, most fluctuating variables depend on  both
of the sample and the temperature used in an experiment. A traditional
way to analyze such systems is introduction of order parameters defined
by replica overlap.~\cite{BY,RSB-review}.

An alternative method~\cite{HA,BrayMoore82}, which 
is more natural and direct, is to choose 
a set of bases adaptive to a given sample 
and temperature. Then, order parameters are defined 
as projections to them. This approach are
also useful for the analysis of Monte Carlo (MC) 
simulation~\cite{Nemoto}.
Recently, Sinova et al.~\cite{Sinova1,Sinova2} pointed out
a relation between the behavior of eigenvalues
of the susceptibility matrix
with the increase of system size and the existence
of replica symmetry breaking (RSB)~\cite{RSB,RSB-review}, which
refreshed the interest in this approach. 
They argue that RSB corresponds
to the existence of \textit{multiple extensive
eigenvalues} in the thermodynamic limit, 
while the droplet picture~\cite{DP} implies
the uniqueness of the extensive eigenvalue. 
Sinova et al. numerically investigated the eigenmodes of the Gaussian
Edwards-Anderson (EA) Ising  model in four dimensions, for which the
existence of RSB is most controversial, and obtained an evidence against
the RSB picture.   

Although their approach is an attractive one, 
their study is limited to relatively small size $\leq 6^4$ (=1296)
of the system. 
In the present letter, we extend the analysis to much larger size
$10^4$ (=10000) and to provide further tests for the RSB scenario. Our
results show that the scaling of the second eigenvalue looks different
in the region of larger system size $\geq 8^4$
at the temperature region that Sinova et al. studied.
It no longer contradicts the RSB picture. 

We also argue that there are no reason 
to restrict ourselves in the analysis of eigenvalues.
Information contained in eigenmodes themselves, 
which are discarded in~\cite{Sinova1,Sinova2}, are
also useful when we consider gauge invariant quantities
defined by them. As an example,
we discuss temperature dependence of the eigenmodes as 
a measure of fragility of the SG equilibrium states.
An evidence of anomalous sensitivity in the SG phase
is obtained by a finite-size-scaling analysis. 

Let us begin with the definition of the model.
The four-dimensional EA model is defined with energy 
\begin{equation}
H(S)=-\sum_{\langle ij\rangle}J_{ij}S_iS_j,
 \label{eqn:Hamilonian}
\end{equation}
where the Ising spins $\{S_i\}$ are defined on a hyper-cubic
lattice in four dimensions with the total number $N=L^4$ of sites.
The strength 
$\{J_{ij}\}$ of nearest-neighbor interactions 
is distributed according to the bimodal distribution 
with equal weights at $J_{ij}=\pm J$. 
A SG phase transition in this model has been well-established by
numerical works. In particular, recent extensive MC studies have
estimated the critical temperature $T_{\rm c}$ to be
$2.0J$~\cite{Marinari,Hukushima}. 
The value of stiffness exponent 
$\theta \sim 0.7$~\cite{Hukushima,Hartmann} in four
dimension is significantly larger than that in three dimensions,
$\theta_{\rm 3D}\sim0.2$~\cite{theta3d}, which
makes the study of asymptotic behavior considerably easier.

The susceptibility matrix of the model
is written as a covariance form
\begin{equation}
C_{ij} = \langle S_iS_j\rangle-\langle S_i\rangle\langle S_j\rangle,
\label{eqn:C}
\end{equation}
where the bracket denotes the thermal average. 
When we perform
a MC simulation for the model, it can be approximated by
$\tilde{C}_{ij}=\frac{1}{M} \sum_{\mu=1}^M X^\mu_i X^\mu_j$, 
where 
$X^\mu_i$ is defined as $S_i^\mu - 1/M \sum_{\mu=1}^M S_i^\mu$
and $S_i^\mu$ is the value of a spin $i \,(i=1\ldots N)$  in a snapshot
$\mu \, (\mu=1 \ldots M)$. 
Then, in principle, diagonalization of $\tilde{C}_{ij}$ gives
eigenvalues and the corresponding eigenvectors of $C_{ij}$.
In the high temperature limit, the matrix ${C}_{ij}$
is equivalent to the interaction matrix $J_{ij}$,
which is sparse in short-range models, but it can be 
a dense matrix at lower temperatures.
An efficient method for the eigenmode analysis 
used in this study will be explained 
at the end of the paper.

In this study, we use the exchange MC method~\cite{EMC}, which
enables equilibration of the system at low temperatures,
even in the SG phase.
We simulate 32 replicas of different values of 
temperature simultaneously with the use of the
multi-spin coding, and try to exchange replicas at neighboring
temperatures after each sweep with single-spin heat-bath flips. 
The lowest temperature of the replicas is $1.0J \sim 0.5T_{\rm c}$, 
whereas the highest temperature is $5.0J \sim 2.5T_{\rm c}$. 
The systems of sizes $L=4$, $6$, $8$, and $10$ are 
examined. For each value of $L$, the number of bond samples used is
$800$, $640$, $464$ and $372$, respectively.

\begin{figure}[]
\resizebox{\figwidth}{!}{\includegraphics{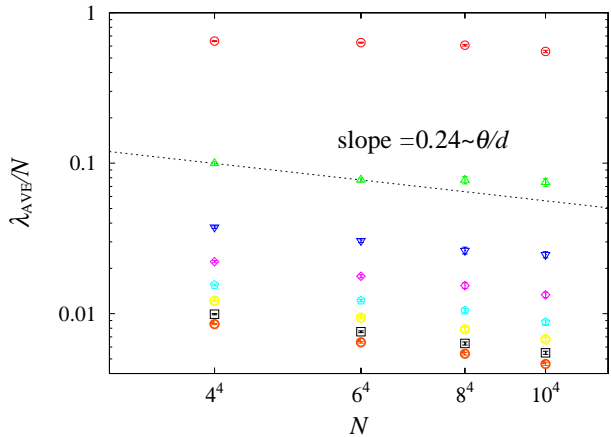}} 
\caption{
Size dependence of the first eight eigenvalues of the susceptibility matrix
 in the four-dimensional EA Ising model at $T/J=1.0$. 
The dotted line has the slope $\theta/d\sim 0.24$. 
}
\label{fig:average}
\end{figure}

In Fig.~\ref{fig:average}, we present the average of
the largest eight eigenvalues $\lambda_{\rm AVE}=[\lambda]_J$ scaled by the system size $N$ with $[\cdots]_J$ being 
an average over the quenched randomness $J_{ij}$.  
The data for the largest scaled eigenvalue is almost independent of $N$,
implying that the eigenvalue is extensive at $T/J=1.0$. 
This is consistent with the assumption that the system is in the SG phase. 
We are mostly interested in whether the second eigenvalue is
extensive or not. It is observed  in Fig.~\ref{fig:average} that the
second eigenvalue apparently follows algebraic decay up to $L \leq 6$, 
which is consistent with the results of Sinova et al.~\cite{Sinova1,Sinova2}. 
The value of the slope 
in $L \leq 6$ in the double-log plot is 
also consistent with 
$\theta/d$ expected from the droplet picture~\cite{Sinova1}, 
when we use the value of $\theta/d$ estimated in the previous
studies~\cite{Hukushima,Hartmann}. 
However, the behavior of the second eigenvalue changes around $L=8$
and has a tendency to saturate to a certain value in $8 \leq L \leq 10$. 
Essentially, the same behavior is observed for  
the typical averaged value, $\lambda_{\rm TYP}\equiv \ln [\exp
\lambda]_J$. 
These observations show that the second eigenvalue is also extensive in
the SG phase. That is,  more than one eigenvalue is extensive. 
According to the interpretation by Sinova et al.\cite{Sinova1,Sinova2},
our results do not contradict the RSB picture. 

Note that our findings of the multiple 
extensive eigenvalues will also be
compatible with a recently proposed sponge-like excitation
picture~\cite{KMPY} (KMPY picture), which argues that
there appear to be 
large-scale low-energy excitations with the fractal dimension $d_s$ less
than the bulk dimension $d$. Since the energy of such an excitation is
supposed not to increase with the system size, we expect that it gives an
additional extensive eigenvalue of the susceptibility matrix.  
The standard RSB and the KMPY pictures can only be distinguished by the
fractal nature of low-lying excitations.

How our results are affected by the critical fluctuation? 
One conventionally uses the overlap distribution function $P(q)$ to 
test the RSB picture in short-range SG
systems~\cite{BY,RSB-review,Reger,Marinari}.  
A non-zero limiting value of $P(q)$ at $q\simeq 0$ in the
thermodynamic limit is considered as an evidence of the RSB phase. 
According to the droplet picture, $P(q)$ at all $q$ except for the
self-overlap \textit{decreases} with increasing $L$ like $L^{-\theta}$. 
On the other hand, at the critical temperature, $P(q=0)$
\textit{increases} with $L$ as $L^{\beta/\nu}$ where 
$\beta$ and $\nu$ are the critical indices. 
These contributions, whose signs are opposite,
can apparently cancel out for moderate system sizes
and give a plateau misidentified as a convergence of $P(0)$
to a non-zero constant~\cite{MBD}. 
Such cancellation would not be expected in the present
approach, because 
both effects make the scaled eigenvalues $\lambda_i/N$ of the
susceptibility matrix \textit{decrease} with
$L$. Thus, there are no chance of the cancellation even in
the critical regions.
We have confirmed that a few largest eigenvalues algebraically
decay in $L$  near the critical temperature, as expected.   
Meanwhile, we see no significant decrease of $\lambda_{\rm AVE}/N$ in
Fig.~\ref{fig:average},  which convinces us that $T/J=1.0$ is already
outside the critical region.

\begin{figure}[]
\resizebox{\figwidth}{!}{\includegraphics{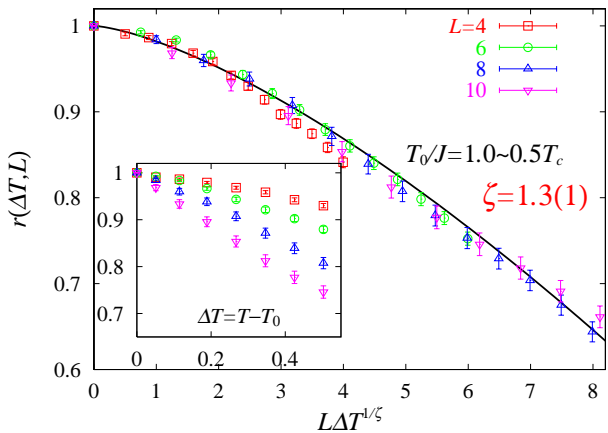}} 
\caption{
A scaling plot of the first eigenmode overlap between $T=T_0=1.0J$ and
 $T_0+\Delta T$ against the scaling variable $L\Delta T^{1/\zeta}$ where
 $\zeta=1.3(1)$. 
The curve represents a fit to the form
 $r(\Delta T,L)=1-CL^{\zeta}\Delta T$~\protect{\cite{fitting}} with $C\sim 0.02$.  
The inset presents raw data as a function of
 temperature difference $\Delta T$ with different sizes. 
} 
\label{fig:tchaos}
\end{figure}

So far we focus our attention on the eigenvalues of the susceptibility
matrix. Useful information is, however, also contained in the eigenmodes. 
Here, as an example,  we propose the use of temperature
dependence of the eigenmodes as a measure of the sensitivity
of the thermodynamic states with respect to a temperature change. 
The overlap between eigenmodes with two different temperatures 
$T_0$ and $T_0+\Delta T$ is defined by the scalar product of
eigenvectors of the susceptibility matrix
\begin{equation}
r(\Delta T,L)=\left[
\left|
\frac{1}{N}\sum_{i} e_i(T_0,L)e_i(T_0+\Delta T,L)
\right|
\right]_J {}_,
\end{equation}
where $e_i(T,L)$ denotes $i$th component of the eigenvector with the
largest eigenvalue. 
We normalize the length of the eigenvectors $e_i$ to the unity. 
With this definition, the overlap $r(\Delta T,L)$ is equal to unity when
the temperature difference $\Delta T$ is zero. 
In the inset of Fig.~\ref{fig:tchaos} we present $r(\Delta T,L)$
calculated with $T_0/J=1.0$. 
For a given temperature difference $\Delta T$, the overlap $r(\Delta
T,L)$ decreases with increasing size $L$. 
We examine an one-parameter scaling $r(\Delta T,L)=R(L/\Delta T^{-1/\zeta})$
for the overlap. As shown in
Fig.~\ref{fig:tchaos}, all the data merges into a
universal scaling function, which is a monotonically decreasing
with the increase of the scaling variable $L\Delta T^{1/\zeta}$.  
The result implies that a pair of eigenmodes
with the largest eigenvalue at infinitesimally different 
temperatures are not
correlated with each other in the thermodynamic limit, 
that is, extreme sensitivity to a temperature perturbation. 
This peculiar feature  never occur in a simple ferromagnet where the
largest eigenmode always corresponds to the uniform state.
It reminds us of ``chaotic nature'' of equilibrium SG
states~\cite{BrayMoore87,chaos_detail}.
It is interesting to point out that
the scaling exponent $\zeta$ defined above
is close to the exponent of chaos associated with 
bond perturbation~\cite{Ney-Nifle}. Whether this coincidence
is accidental or not is left for future studies. 

Let us give a comment on the numerical method for
calculation of the eigenmodes, which relaxes the
numerical difficulty of treating a covariance matrix of large
dimensions. In this method, \textit{dual algorithm},  we
diagonalize a  $M \times M$ matrix $\tilde{D}$ 
defined by
\begin{equation}
\tilde{D}_{\mu\nu}= \frac{1}{N} \sum_{i=1}^N X^\mu_i X^\nu_i, 
\end{equation}
instead of diagonalizing the $N \times N$ matrix $\tilde{C}$. 
We denote eigenvectors of the matrices $\tilde{C}$ 
and $\tilde{D}$ 
as $\{e_i\}^{(n)}$ and $\{e^\mu\}^{(m)}$, respectively, where
$n$ and $m$ are the indices of eigenvectors. 
It is easy to show the following primal--dual relations:  
(1)~$\sum_i X_i^\mu e_i^{(n)}$ is an eigenvector of 
$\tilde{D}$ if it is not the null vector, and 
(2)~$\sum_\mu X_i^\mu {e^\mu}^{(m)}$ is an eigenvector 
of $\tilde{C}$ if it is not the null vector.  
Note that when $N>M$, the matrix 
$\tilde{C}$ is singular and has at least $N-M$
zero eigenvalues, and, conversely, when $M>N$, 
the matrix $\tilde{D}$ is
singular and has at least $M-N$ zero eigenvalues. 
With this relation, we can replace the diagonalization of the covariance
matrix  $\tilde{C}$ with that of the sample-overlap matrix $\tilde{D}$, 
which reduces the amount of computation when $N>M$.
Both algorithms give the same results for the same set of samples 
within numerical accuracy.
We should use a sufficient large number $M$ which gives a good approximation
of the susceptibility matrix. 
In the present study, we have tested convergence of the eigenvalues with
different  values of $M$ and have confirmed that the algorithm shows
good  convergence. 
The dual algorithm has been known in the field of multivariate
analysis  and already used in simulation studies of proteins. 
In the field of spin glasses, however, it seems less known and used. 
While  a dual plot for visualizing hierarchical structures of low-energy
valleys is recently introduced  by \cite{Domany01}, they did not stress
the advantage of the dual formulation as a tool for efficient
computation.  

To summarize, we numerically explored the eigenmodes of the
susceptibility matrix of the four-dimensional Edwards-Anderson
Ising spin glass model.
First, we studied the eigenvalues of systems larger
than those investigated in the previous study and 
found a strong evidence that the second eigenvalue is extensive in the
thermodynamic limit. This no longer conflicts with the RSB pictures. 
Secondly, we discuss the sensitivity of the normalized 
eigenmodes against temperature perturbation and found extreme
sensitivity to the variation of the temperature. Finally, we mention to a
technique with the primal--dual relation used in this study, 
which will be useful for the study of large systems. 

\begin{acknowledgments}
We would like to thank K.~Nemoto, who kindly introduced us to this topic
 and gave us a number of important suggestions. 
We are grateful to H.~Yoshino for suggesting the functional form of
 scaling in Fig.~\ref{fig:tchaos}. We also thank H.~Takayama for fruitful
 discussions and continuous encouragement. 
K.H. acknowledges a Grant-in-Aid for the Encouragement of
 Young Scientists (\#13740233) from the Ministry of Education, Culture,
 Sports, Science and Technology of Japan. The present simulations have
 been performed on SGI 2800/384 at the Supercomputer Center, Institute
 for Solid State Physics, the University of Tokyo. 
\end{acknowledgments}


\begin{thebibliography}{99}
\bibitem{chaos_detail}
The chaotic nature has been established for a change of the random
bond $J_{ij}$ and magnetic field~\cite{Ritort94,Ney-Nifle}, while the
chaos arising from the temperature change still remains very
controversial~\cite{BilloireMarinari,KrzakalaMartin}.
\bibitem{fitting}
According to the scaling argument~\protect{\cite{BrayMoore87,DP}}, 
the probability of large scale excitations induced by the perturbation
	 is given by $\Delta TL^\zeta$. 
It leads to this fitting formula, if the function
	 $1-r(\Delta T,L)$ is determined by the probability of the large
	 scale excitations.
\end{thebibliography}
\end{document}